\documentclass[aps,prl,twocolumn,floatfix,nofootinbib,showpacs,superscriptaddress]{revtex4-1}
\usepackage{amsmath,amsfonts,amssymb,bm}
\usepackage{graphicx}
\usepackage{subfigure}
\usepackage{color}
\definecolor{purple}{rgb}{0.5,0,0.5}
\definecolor{blue}{rgb}{0.0,0,0.9}
\usepackage[colorlinks=true, pdfstartview=FitV, linkcolor=purple, citecolor= purple, urlcolor=blue]{hyperref}

\begin{document}

\title{Colour Confinement: a Dynamical Phenomenon of QCD}

\author{Fei Gao }
\affiliation{Department of Physics and State Key Laboratory of Nuclear Physics and Technology,
Peking University, Beijing 100871, China}
\affiliation{Collaborative Innovation Center of Quantum Matter, Beijing 100871, China}

\author{Chong-yao Chen }
\affiliation{Department of Physics and State Key Laboratory of Nuclear Physics and Technology,
Peking University, Beijing 100871, China}
\affiliation{Collaborative Innovation Center of Quantum Matter, Beijing 100871, China}

\author{Yu-xin Liu }
\email[Corresponding author: ]{yxliu@pku.edu.cn}
\affiliation{Department of Physics and State Key Laboratory of Nuclear Physics and Technology,
Peking University, Beijing 100871, China}
\affiliation{Collaborative Innovation Center of Quantum Matter, Beijing 100871, China}
\affiliation{Center for High Energy Physics, Peking University, Beijing 100871, China}

\date{\today}

\begin{abstract}
We study in this Letter the origin of the confinement in QCD by analyzing the colour charge of physics states.
We derive the colour charge operator in QCD and compare it with the electromagnetic charge operator in QED.
It shows that the two charges have very similar structure, but the dynamical properties of the gauge fields are different. The difference between the behaviours of the gauge boson propagator at zero momentum for QCD and that for QED guarantees that there occurs colour confinement in QCD but there is no confinement in QED.
We give then a universal relation between the confinement and the dynamical property of QCD and reveals the origin of the colour confinement, which can be demonstrated as the dynamical effect of QCD
or more explicitly the dynamical mass generation of the gluon.
\end{abstract}


\maketitle

\noindent{\emph{Introduction:}}--- Even though quarks and gluons have been confirmed as the elementary fields of Quantum Chromodynamics (QCD), they cannot be observed in physics space.
Such a bizarre phenomenon in QCD is called colour confinement.
People have made many efforts on understanding the confinement. Some people have studied the static quark potential, and considered the linear rise behaviour as confinement. The linear potential can be obtained by the colour flux-tube  model~\cite{Kogut:1976NPB,Buchmuller:1982PLB,Isgur:1985PRD}, Casimir effect~\cite{Faber:1998PRD,Bali:2000PRD}, lattice QCD simulations\cite{Kogut:1983RMP,Takahashi:2002PRD}, related to the Gribov horizon~\cite{Zwanziger:2003PRL,Greensite:2004PRD} and Light-Front Holography~\cite{Brodsky:2017}.
However, the potential description is lack of direct connection with QCD and mostly depends on numerical calculations.
Some people have considered the definite-positivity violation of the propagator's spectral density of the particles 
as the criterion labeling the confinement
%
%
(see, e.g., Refs.~\cite{Roberts:1994PPNP,Alkofer:2000PR,Qin:2013PRD,Gao:2016PRD1}) or the zero-mode of the spectral density function~\cite{Qin:2011PRD,Gao:2014PRD,Su:2015PRL,Zwanziger:2016PRD}.
Other people compute some specific configurations of gauge fields, such as instantons, merons and center vortices, to explain the confinement in QCD with the Z(3) center symmetry or dual superconductivity based on lattice QCD simulation
(see, e.g., Refs.~\cite{Cornwall:1979NPB,Teper:1985PLB,Greensite:2003PPNP,Kondo:2008PLB}) and the modifications~\cite{Oxman:2013JHEP,Kharzeev:2015PRL,Dudal:2016PRD,Alkofer:2018PLB},
or the Borromean picture~\cite{Roberts:2015PLB}.
Such investigations give strong hints on the confinement, but are still not essential enough to reveal the origin of the confinement.
Gribov-Zwanziger/Kugo-Ojima scenarios~\cite{Gribov:1978NPB,Zwanziger:1998NPB,Kugo:1979PTPs,Hata:1982PTP} refer the colour charge to the Becchi-Rouet-Stora-Tyutin  (BRST) charge and show that an infrared enhancement of the ghost and an infrared suppression of the gluon signal the confinement.
However, it has not yet reached a clear origin for the  confinement,
since the operator with BRST charge is just a special treatment of the gauge fixing and is irrelevant to the physics states.
In addition, succeeded numerical calculations provide evidence that the confinement requires a finite gluon mass (see, e.g., Refs.~\cite{Pawlowski:2004PRL,Pawlowski:2010PLB,Pawlowski:2012PRD}) and very recent general analysis indicates that the ghost should be infrared constant~\cite{Gao:2017PLB}.
The scheme also has difficulty on distinguishing the differences between QCD and Quantum Electrodynamics (QED),  and is thus not satisfactory sufficiently.
Even though there have been some other attempts, such as, the hidden local symmetry~\cite{Kitano:2011JHEP},
the SU(N) Euclidean Yang-Mills theory~\cite{Matsuoka:2012PRD},
the supersymmetric quiver gauge theory~\cite{Kitano:2013JHEP}, and so on, 
until now there is still no pronounced analytic proof for why colour confinement exists in QCD.

To understand the origin of the colour confinement, one needs to recognize the differences  between QCD and QED.  Generally, there're two  features that QCD differs from QED,
namely, QCD is a non-Abelian gauge theory and strong coupling theory.
If confinement is owing to the non-Abelian nature of the theory essentially but not concerning about the coupling strength, the perturbative calculation should have had shown some hints.
The opposite fact indicates that the confinement is somehow connected with the strong coupling of QCD. However, the strong coupling property also becomes an obstacle to understand the confinement in QCD since it is very difficult to carry out direct and complete calculation.
If just employing the nonperturbative numerical methods or some effective models,
the results are model-dependent and it is very hard to demonstrate the origin of the confinement.
An analytic proof for the existence of the confinement in QCD is then referred to a millennium problem.

In this Letter, we find a universal relation between the confinement and the dynamical property of the gauge fields by analyzing the structures of the electromagnetic charge and the colour charge operators carefully.
We observe that these two charge operators have quite similar structure.
The key difference between them is a factor that is related to the dynamical property of the gauge field's propagator at zero momentum.
For QED, there's a massless pole in the photon propagator,
the electromagnetic charge becomes nonzero with this singularity.
Meanwhile, it has been known that there is no such kind pole in the gluon propagator,
it leads then to  vanishing colour charges for any physics states.

\medskip

\noindent{\emph{Derivations and Discussions:}}---
At first, we retrieve the case of QED. The electromagnetic current can be defined straightforwardly as the Noether current
$$ J_{\mu}^{} = \sum_{i} \frac{\partial{\mathcal{L}}}{\partial{\partial_{\mu}^{}\phi_{i}^{} }}\Delta\phi_{i}^{} $$ %
with $\Delta\phi_{i}^{}$ the global gauge transformation that keeps the Lagrangian ${\mathcal{L}}$ invariant (or equivalently, the local transformation that keeps the action invariant), which reads explicitly
\begin{equation}
J^{e}_{\mu}(x) = e \bar{\psi}(x) \gamma_{\mu}^{} \psi(x) \, .
\end{equation}
The equation of motion is
$$ e \bar{\psi}(x) \gamma_{\mu}^{} \psi(x) = \partial_{\nu}^{} F_{\nu\mu}^{}(x) $$
with gauge fixing condition
$$ \big{\langle} \mathrm{phys} \big{|} \partial_{\mu}^{} A_{\mu}^{} \big{|} \mathrm{phys} \big{\rangle} = 0 \, . $$
The gauge tensor of QED is
$$ F_{\mu\nu}^{} = \partial_{\mu}^{} A_{\nu}^{} - \partial_{\nu}^{} A_{\mu}^{} \, . $$
With the electromagnetic charge operator defined as
$$ \hat{Q}_{e}^{}= \int d^{3} x \partial_{\mu}^{} F_{\mu0}^{} \, , $$
we have the expectation value of the electromagnetic charge
\begin{eqnarray} \label{eq:qed}
\begin{split}
   & \big{\langle} \mathrm{phys} \big{|} \hat{Q}_{e}^{} \big{|} \mathrm{phys} \big{\rangle}
=  \big{\langle} 0 \big{|} a^{r}_{\vec{q}} \hat{Q}_{e}^{} a^{s\dagger}_{\vec{p}} \big{|} 0 \big{\rangle} \\
= & \Big{\langle} 0 \Big{|} a^{r}_{\vec{q}} \int d^{3} x  \big{(} \partial^{2} A^{0} (x) - \partial^{0} \partial_{\mu}^{} A_{\mu}^{} (x) \big{)} a^{s\dagger}_{\vec{p}} \Big{|} 0 \Big{\rangle} \, ,
\end{split}
\end{eqnarray}
where $a^{s\dagger}_{\vec{p}}$ and $a^{r}_{\vec{q}}$ are, respectively, the generation, annihilation operator of the physics state (on-shell state) in the Fock representation, which reads
\begin{subequations}
\begin{eqnarray}
\psi(x) & = & \int \frac{d^{3} \vec{p}}{(2\pi)^3}\frac{1}{\sqrt{2E_{\vec{p}}^{}}}
    \sum_{s=1,2} \Big{(} a^{(s)}_{\vec{p}} u^{(s)}(p) e^{-ip\cdot x}       \notag\\
& & \qquad
           + \, b^{(s)\dagger}_{\vec{p}} v^{(s)}(p) e^{ip\cdot x} \Big{)} \, ,  \\
\bar{\psi}(x) & = & \int \frac{d^{3} \vec{p}}{(2\pi)^3} \frac{1}{\sqrt{2E_{\vec{p}}^{} } }
     \sum_{s=1,2} \Big{(}  a^{(s)\dagger}_{\vec{p}} \bar{u}^{(s)}(p) e^{ip\cdot x}     \notag \\
& & \qquad
           + \, b^{(s)}_{\vec{p}} \bar{v}^{(s)}(p) e^{-ip\cdot x} \Big{)}  \, ,
\end{eqnarray}
\end{subequations}
where $E_{\vec{p}}^{}= p_{0}^{}$, $b^{(s)}_{\vec{p}}$,$b^{(s)\dagger}_{\vec{p}}$ are the generation, annihilation operator of anti-particles,
$u^{(s)}(p)$, $\bar{u}^{(s)}(p)$, and $v^{(s)}(p), \; \bar{v}^{(s)}(p)$ are the Dirac spinors.
With the gauge fixing condition, the second term in the r.h.s. of Eq.~(\ref{eq:qed}) vanishes.
After contracting the operators, we get
\begin{eqnarray}
&& \big{\langle} 0 \big{|} a^{r}_{\vec{q}} \hat{Q}_{e}^{} a^{s\dagger}_{\vec{p}} \big{|} 0 \big{\rangle} \notag\\
& = & \int d^{3} x \int d^{4} y \int d^{4} k \frac{-e}{\sqrt{4E_{\vec{p}}^{} E_{\vec{q}}^{}}} k^{2} D(k^{2}) \notag \\
& & \times e^{ik(x-y)} e^{i(p-q)y} tr[\gamma_{0}^{} u^{s}(p) \bar{u}^{r}(q) ] \notag \\
& = & \int d^{4}k \frac{-e}{2\sqrt{E_{\vec{p}}^{} E_{\vec{q}}^{} }} k^{2} D(k^{2}) \notag \\
&& \times \delta^{4}(k-p+q) \delta^{3}(\vec{k}) e^{ik_{0}^{} x_{0}^{}}
     tr[\gamma_{0}^{} u^{s}(p) \bar{u}^{r}(q)] \, ,
\end{eqnarray}
with the photon propagator defined as $-i\eta_{\mu\nu}^{} D(k^{2})$ and $\eta_{\mu \nu}^{}$ the metric tensor.
Since the momentum $p$ and $q$ are the on-shell fermion's momentum,
the $\delta$-functions actually lead to $p=q$, and thus, the charge becomes
\begin{equation}
\big{\langle} 0 \big{|} a^{r}_{\vec{q}} \hat{Q}_{e}^{} a^{s\dagger}_{\vec{p}} \big{|} 0 \big{\rangle}
= - e\delta^{rs} \delta^{3}(\vec{p} \! - \! \vec{q}) \! \int \! d^{4}k k^{2} D(k^{2}) \delta^{4}(k) \, .
\end{equation}
For the electron in QED, since the photon propagator holds $D(k^2)=1/k^{2}$,
the relation becomes
$$ \big{\langle} 0 \big{|} a^{r}_{\vec{q}} \hat{Q}_{e}^{} a^{s\dagger}_{\vec{p}} \big{|} 0 \big{\rangle} = - e \big{\langle} 0 \big{|} a^{r}_{\vec{q}} a^{s\dagger}_{\vec{p}} \big{|} 0 \big{\rangle} = - e \delta^{r s} \delta^{3}(\vec{p} \! - \! \vec{q}) \, , $$
which means that the electron carries one unit negative electromagnetic charge.
Recalling the analyzing process, one can notice that the integration in the charge operator is directly a volume integration on the $\partial_{\mu}^{} F_{\mu\nu}^{}$,  or actually a surface integration on the $F_{\mu\nu}^{}$.
If there is no singularity inside the surface, the integration must vanish due to the antisymmetric nature of the $F_{\mu\nu}^{}$.
For QED, the states are not always zero charge since there exists singularity in the gauge tensor.
This indicates distinctly that the massless pole of the photon propagator is essential to the nonvanishing charge.
More theoretically speaking, the singularity at zero momentum in the photon propagator contributes to the term $\partial_{\mu}^{} F_{\mu\nu}^{}$ and leads to a finite value.

Now we go to QCD. We employ at first a similar gauge fixing procedure with that in QED for the convenience of comparison.
With the Faddeev-Popov procedure~\cite{Faddeev:1967PLB}, we will certainly get the same results
that will be discussed latter.
Analog to QED, we take the gauge invariant action:
\begin{eqnarray}
S&=&\int d^4x \mathcal{L}_{QCD} \, ,   \\
\mathcal{L}_{QCD}^{} & = & -\frac{1}{4}F^{a}_{\mu\nu} F^{a,\mu\nu} \!
 + \bar{\psi} (i\partial\!\!\!/ \! - \! m)\psi + g A^{a}_{\mu} \bar{\psi} \gamma_{\mu}^{} t^{a} \psi \, ,  \qquad
\end{eqnarray}
where
$$F^{a}_{\mu\nu} = \partial_{\mu}^{} A^{a}_{\nu} - \partial_{\nu}^{} A^{a}_{\mu}
+ gf^{abc} A^{b}_{\mu} A^{c}_{\nu}$$
with $g$ the coupling constant, $A^{a}_{\mu}$ the gauge boson field and $f^{abc}$ the structure constant of the gauge group.
$\bar{\psi}$, $\psi$  is the anti-quark, quark field, respectively,
and  $t^{a}$  the generators of the gauge group.
The gauge fixing condition is defined as
$$\big{\langle} \mathrm{phys} \big{|} \partial_{\mu}^{} A^{a}_{\mu} \big{|} \mathrm{phys} \big{\rangle} =0 \, , $$
where the physics state, $\vert  \mathrm{phys} \rangle$, means any on-shell state.
The colour gauge current can be defined based on the global colour transformation (with SU(3) symmetry)
in the similar way as that in QED , which reads
\begin{equation}
J^{a}_{\mu} (x) = g \bar{\psi}(x) \gamma_{\mu}^{} t^{a} \psi(x) + g f^{abc} A^{b}_{\nu}(x) F^{c}_{\nu\mu}(x) \, ,
\end{equation}
and the colour charge operator is
$$ \hat{Q}_{c}^{a}=\int d^{3} x J^{a}_{0} (x) \, . $$
%
%
The equation of motion of the gluon is given by the variation of the field, which can be written explicitly as
\begin{equation}
J^{a}_{\mu}(x) = \partial_{\nu}^{} F^{a}_{\nu\mu}(x) \, .
\end{equation}
The colour charge of the physics states can then be converted into
\begin{equation}
q_{A}^{}  = \big{\langle} \mathrm{phys} \big{|} \int d^{3} x \partial_{\nu}^{} F^{a}_{\nu0}(x) \big{|} \mathrm{phys} \big{\rangle} \, .
\end{equation}
%
%
In more detail, the colour charge can be expanded as:
\begin{eqnarray}   \label{eqn:cc1}
\begin{split}
& \Big{\langle} \mathrm{phys} \Big{|} \hat{Q}_{c}^{a} \Big{|} \mathrm{phys} \Big{\rangle} \\
= \, & \Big{\langle} \mathrm{phys} \Big{|} \int d^{3} x \Big{\{ } \partial^{2} A^{a}_{0}(x)
- \partial^{0} \partial_{\mu}^{} A^{a}_{\mu}(x)   \\
& \quad + g f^{abc} \partial_{\mu}^{} (A^{b}_{\mu}(x) A^{c}_{0} (x)) \Big{\}} \Big{|} \mathrm{phys} \Big{\rangle}  \, .
\end{split}
\end{eqnarray}

Similar to that in QED, the second term in the r.h.s of Eq.~(\ref{eqn:cc1}) vanishes since we employ the same gauge fixing condition.
Recalling that the interaction terms attached to the field $A_\mu^a(x)$  in the action are $\int d^{4}x A_{\mu}^{a} (x) \big(J^{a}_{\mu}(x) - \partial_{\nu}^{} F^{a}_{\nu\mu}(x)\big)$ , one can expand the  term $\partial^{2} A^{a}_{0}(x)$ in terms of the interaction operators (similarly expressed as: $ \langle A_{\mu}^{}(x) \rangle = \int d^{4}y D_{\mu\nu}^{}(x-y) \langle j_{\nu}^{} (y) \rangle $ with
$S_{int}^{} = \int d^{4}x A_{\mu}^{}  j_{\mu}^{}$).
Considering further the gluon propagator  $D_{\mu\nu}^{}(x-y)=\int d^{4} k \eta_{\mu\nu}^{}D(k^2)e^{ik(x-y)}$ with $\eta_{\mu \nu}^{}$ the metric tensor,
one can rewrite the colour charge as
\begin{eqnarray}   \label{eqn:cc2}
\begin{split}
  & \Big{\langle} \mathrm{phys} \Big{|} \hat{Q}_{c}^{a} - \int d^{3} x g f^{abc} \partial_{\mu}^{}
\big{(} A^{b}_{\mu}(x) A^{c}_{0} (x) \big{)} \Big{|} \mathrm{phys} \Big{\rangle} \\
=\; &  \int d^3x \int d^{4} k k^{2} D(k^{2}) \int d^{4} y e^{ik_{}^{} (x_{}^{} - y_{}^{})} \\
 & \; \times \Big{\{} \Big{\langle} \mathrm{phys} \Big{|} J^{a}_{0} (y) \Big{|} \mathrm{phys} \Big{\rangle} \\
 & \qquad - \Big{\langle} \mathrm{phys} \Big{|} g f^{abc} \partial_{\mu}^{} (A^{b}_{\mu}(y) A^{c}_{0}(y)) \Big{|} \mathrm{phys} \Big{\rangle} \Big{\}} \\
=\; &  \int d^{4} k k^{2} D(k^{2}) \delta^{3}(\vec{k}) \int d^{4} y e^{ik_{0}^{} (x_{0}^{} - y_{0}^{})} \\
 & \; \times \Big{\{} \Big{\langle} \mathrm{phys} \Big{|} J^{a}_{0} (y) \Big{|} \mathrm{phys} \Big{\rangle} \\
 & \qquad - \Big{\langle} \mathrm{phys} \Big{|} g f^{abc} \partial_{\mu}^{} (A^{b}_{\mu}(y) A^{c}_{0}(y)) \Big{|} \mathrm{phys} \Big{\rangle} \Big{\}} \, .
\end{split}
\end{eqnarray}

Since the momentum $p$, $q$ and the Lorentz indices of the in-state and out-state are symmetric in the above definition,
the term $\partial_\mu \big(f^{abc}A^b_\mu(x) A^c_\nu(x)\big)$ cancels if we look into the contraction of the operators.
This can be understood through analysing the Lorentz structure,
 $f^{abc} A^{b}_{\mu}(x) A^{c}_{\nu}(x)$ can only be  proportional to $\eta_{\mu\nu}^{}$ or $p_{\mu} p_{\nu}$,  and thus this term vanishes. We then get:
\begin{eqnarray}   \label{eqn:cc3}
\begin{split}
 & \Big{\langle} \mathrm{phys} \Big{|} \hat{Q}_{c}^{a}  \Big{|} \mathrm{phys} \Big{\rangle} \\
=\; &  \int d^{4} k k^{2} D(k^{2}) \delta^{3}(\vec{k}) \int d^{4} y e^{ik_{0}^{} (x_{0}^{} - y_{0}^{})} \\
 & \; \times 
 \Big{\langle} \mathrm{phys} \Big{|} J^{a}_{0} (y) \Big{|} \mathrm{phys} \Big{\rangle}   
 \, .
\end{split}
\end{eqnarray}

Noticing in Eq.~(\ref{eqn:cc3}) the charge operator nature of  $\hat{Q}_{c}^{a}$ and the inner product in the r.h.s., one can recognize that the expectation value for each can not depend on time $x_{0}^{}$ and $y_{0}^{}$, respectively. This is simply just the manifestation of the momentum conservation with $k=p-q=0$.   Then the integration $\int dy_{0}^{} e^{ik_{0}^{}(x_{0}^{} - y_{0}^{} )}$ in the r.h.s. of Eq.~(\ref{eqn:cc3}) contributes a term $\delta(k_{0}^{})$. The above expectation value can thus be rewritten as
\begin{equation}
\Big{\{} 1 -\int d^{4} k \delta^{4}(k) k^{2} D(k^{2} ) \Big{\}}
\Big{\langle} \mathrm{phys} \Big{|} \hat{Q}_{c}^{a} \Big{|} \mathrm{phys} \Big{\rangle} = 0 \,.
\end{equation}

This is a universal relation between the colour charge and the dynamical property of QCD.
It indicates that the confinement is actually owing to the dynamical effect of QCD.
If the gluon propagator does not carry a massless pole with  residue one,
i.e., $\int d^{4} k \delta^4(k) k^{2} D(k^{2}) \neq 1 $,
the colour charge is then zero for any physics state.
In other word, the confinement demands a nonzero mass scale generated dynamically to eliminate the massless pole from the gauge boson propagator.
Functional renormalization group calculation and Dyson-Schwinger equation computations (e.g., Refs.~\cite{Pawlowski:2010PLB,Aguilar:2012PRD,Ayala:2012PRD,Strauss:2012PRL}), lattice QCD simulations (e.g., Refs.~\cite{Bogolubsky:2009PLB,Boucaud:2010PRD,Bowman:2004PRD,Cucchieri:2012PRD,Oliveira:2011JPG,Pawlowski:2012PRD})  and general analyses (e.g., Refs.~\cite{Qin:2011PRC,Zwanziger:2013PRD}) have approved that there really exists a dynamically generated non-zero mass scale,
and there is no massless pole at zero momentum for the gluon propagator.
Such dynamical behaviours reveal then the origin of the colour confinement in QCD.
It is now clear that the confinement is equivalent to the dynamical mass generation of the gluon.

If we employ the local covariant Faddeev-Popov formalism of the QCD action,
the formula for the colour charge and the equation of motion becomes different
since the ghost and anti-ghost fields are added.
The global gauge transformation that keeps the Lagrangian invariant now becomes \cite{Kugo:1979PTPs}:
\begin{eqnarray}
\begin{split}
\delta\psi = - ig \theta^{a} t^{a} \psi, \qquad & \delta A^{a}_{\mu} = gf^{abc}\theta^{b} A^{c}_{\mu} , \qquad \\
\delta c^{a} = gf^{abc} \theta^{b} c^{c} , \qquad & \delta \bar{c}^{a}=g f^{abc} \theta^{b} \bar{c}^{c} . \qquad
\end{split}
\end{eqnarray}
where  $c^{a}$ and $\bar{c}^{a}$ are the ghost, anti-ghost fields with $\theta^{i}$ the phase angle.

The colour current can be expressed as
\begin{eqnarray}
\begin{split}
 J_{\mu}^{a} = &\; g \bar{\psi}(x) \gamma_{\mu}^{} t^{a} \psi(x) + gf^{abc} A^{b}_{\nu} (x) F^{c}_{\nu\mu}(x) \\
       & - \frac{1}{\xi} f^{abc} A^{b}_{\mu}(x) \partial_{\nu}^{} A^{c}_{\nu}(x) \\
       & + g f^{dba} f^{dec} \bar{c}^{b} (x) A_{\mu}^{c} (x) c^{e}(x) \, ,
\end{split}
\end{eqnarray}
where $\xi$ is the gauge parameter.
Operating on physics states, the last term contributes colour structure $f^{dba}f^{dec}f^{bce}$, and thus vanishes.
The effective colour charge can then be written as
$$ \Big{\langle} \mathrm{phys} \Big{|} \hat{Q}_{c}^{a} \Big{|} \mathrm{phys} \Big{\rangle}
= \Big{\langle} \mathrm{phys} \Big{|} \int d^{3} x \bar{\psi}(x) \gamma_{0} t^{a} \psi(x) \Big{|} \mathrm{phys} \Big{\rangle} \, . $$
The equation of motion becomes
\begin{eqnarray}
\begin{split}
    & g\bar{\psi}(x) \gamma_{\mu}^{} t^{a} \psi(x)
     + g f^{abc} A^{b}_{\nu}(x) F^{c}_{\nu\mu}(x)     \\
    & \;\;  + g f^{abc}\partial_{\nu}^{} \bar{c}^{b}(x) c^{c}(x) \\
 = & \; \partial_{\nu}^{} F^{a}_{\nu\mu}(x) + \frac{1}{\xi}\partial_{\mu}^{} \partial_{\nu}^{} A^{a}_{\nu}(x) \, .
\end{split}
\end{eqnarray}

With the same procedure as we take in case that the colour transformation is taken in the way similar to that in QED, we get the relation
\begin{eqnarray}
\begin{split}
  & \Big{\{} 1 \! - \! \int d^{4} k \delta^{4}(k) \eta_{\mu0}^{} \Big{(} \eta_{\mu\nu}^{} k^{2} \! - \!
      \Big{(}1 \! - \! \frac{1}{\xi} \Big{)} k_{\mu}^{} k_{\nu} \Big{)} D_{\nu\mu^{\prime}}^{}(k^{2}) \Big{\}} \qquad \\
  & \times \int d^{3} x \Big{\{} \Big{\langle} \mathrm{phys} \Big{|} g f^{abc} D_{\mu^{\prime}}^{}
      \bar{c}^{b} (x) c^{c} (x)  \Big{|} \mathrm{phys} \Big{\rangle}     \\
  & \qquad - \frac{1}{\xi} \Big{\langle} \mathrm{phys} \Big{|}  f^{abc} A^{b}_{\mu^{\prime}}
      \partial_{\nu^{\prime}}^{}  A^{c}_{\nu^{\prime}}(x) \Big{|} \mathrm{phys} \Big{\rangle}  \\
  & \qquad + \Big{\langle} \mathrm{phys} \Big{|}  J^{a}_{\mu^{\prime}}(x) \Big{|} \mathrm{phys} \Big{\rangle} \Big{\}} = 0 \, ,
\end{split}
\end{eqnarray}
where $D_{\mu^{\prime}}^{}\bar{c}^{a}(x) = \partial_{\mu^{\prime}}^{} \bar{c}^{a}(x) 
- g f^{abc} A_{\mu^{\prime}}^{b} \bar{c}^{c}(x)$.
The first two expectation values on physics states come from the gauge fixing term in the Faddeev-Popov procedure to eliminate the unphysical component in the physics states.
It is just the  Slavnov-Taylor identity of the ghost field from the BRST symmetry,
it thus vanishes for all physics states.
With the gluon propagator defined as
$$ D_{\mu\nu}^{}(k^{2}) = \Big{(} \eta_{\mu\nu}^{} - ( 1 - \xi ) \frac{k_{\mu}^{} k_{\nu}^{}}{k^{2}} \Big{)}  D(k^{2}) \, , $$
we get the same relation as that obtained above,
$$
\Big{\{} 1 - \int d^{4} k \delta^{4}(k) k^{2} D(k^{2}) \Big{\}} \Big{\langle} \mathrm{phys} \Big{|}
\hat{Q}_{c}^{a}  \Big{|} \mathrm{phys} \Big{\rangle} = 0 \, .
$$
This manifests apparently that the results with different gauge fixing schemes are certainly equivalent.

It is also interesting to investigate the Higgs mechanism where the mass of the SU(2) gauge boson  are generated spontaneously via the coupling with the Higgs boson.
Because the equation of motion involves the Higgs boson,
there is an additional interaction term: $g^{2} \phi^{2} (x) A^{i}_{\mu}(x)$ with $\phi(x)$ the Higgs field
and $A^{i}_{\mu}(x)$ the gauge boson field.
This term modifies the charge operator's expectation value to $\int d^{4} k \delta^{4}(k)
\big{(} k^{2} - g^{2} \langle \phi^{2} \rangle \big{)} D(k^{2}) \big{\langle} \mathrm{phys} \big{|} \hat{Q}_{A}^{} \big{|} \mathrm{phys} \big{\rangle} $.
Therefore, no matter whether the symmetry is spontaneously broken or not,
the coefficient of this term keeps being one and then the charge structure is the same as that in QED.
Consequently, there is certainly no confinement in the theory.

In addition, it is necessary to explore whether there is confinement when the coupling constant of QED is strong enough.
The answer is probably no.
In Ref.~\cite{Gao:2018PRD}, it shows that the gluon mass is supposed to have strong connection with the Gribov horizon,
which is a special property of the SU(N) gauge theory.
This indicates that in QED which has no Gribov horizon,
the dynamical effect might not generate a non-zero mass for the photon at zero momentum,
and thus there does not exit confinement.

\medskip

\noindent{\emph{Summary:}}---We show in this Letter that the colour charge and electromagnetic charge operators have very similar structure. The charges can be converted into an integral with the dynamical behaviours of the corresponding gauge fields.
Such an integral is in fact a surface integration, and vanishes if there is no singularity inside.
More concretely, there is a factor $k^{2}D(k^{2})\delta^{4}(k)$, with $D(k^{2})$ being the propagator of the gauge particle, in the integrand of the expectation value of the charge operator,
which reveals the relation between the confinement and the dynamical property of the theory essentially. %
For QED, the massless pole of the $D(k^{2})$ at zero momentum in the photon propagator ($1/k^{2}$)
leads to nonzero electromagnetic charge, for instance, the charge is $-e$ for electron.
While for QCD, the dynamical effect makes the gluon have nonzero mass.
The fact that there is no massless singularities in the gauge tensor guarantees zero colour charge for any physics state, in turn, there exists colour confinement.
In addition, recent numerical results have confirmed that the gluon does not have the similar pole at zero momentum and they are more supportive of an infrared-finite gluon propagator.
The consistence of recent numerical results that there is a nonzero mass scale generated dynamically for gluon with the condition for  the colour confinement may have provided the evidence for our mechanism of the colour confinement.

\medskip

The work was supported by the National Natural Science Foundation of China under Contracts No. 11435001 and
No. 11775041; the National Key Basic Research Program of China under Contracts No.~2013CB834400
and No.~2015CB856900.

\end{document}